\documentclass[aps,twocolumn,showpacs,citeautoscript,reprint]{revtex4-2}

\usepackage{bm}
\usepackage{graphicx}
\usepackage{amsmath}
\usepackage{amsfonts}
\usepackage{siunitx}
\usepackage{bbold}
\usepackage{color}
\usepackage{comment}
\usepackage{textcomp}
\usepackage{gensymb}
\usepackage{soul}

\begin{document}

\title{Impact of vacancies on 
twisted bilayer graphene quantum point contacts}%

\author{Pablo Moles}

\affiliation{GISC, Departamento de F\'{\i}sica de Materiales, Universidad Complutense, E-28040 Madrid, Spain}

\author{Francisco Dom\'{\i}nguez-Adame}

\affiliation{GISC, Departamento de F\'{\i}sica de Materiales, Universidad Complutense, E-28040 Madrid, Spain}

\author{Leonor Chico}

\affiliation{GISC, Departamento de F\'{\i}sica de Materiales, Universidad Complutense, E-28040 Madrid, Spain}

\date{\today}
\begin{abstract}

We carry out an extensive numerical study of low-temperature electronic transport in quantum point contacts based on twisted bilayer graphene. Assuming ballistic electron dynamics, quantized plateaus in the conductance are observed in defect-free samples when the twisting angle is large enough. However, plateaus are smeared out and hardly noticeable on decreasing the angle. Close to the magic angle, the conductance around the charge neutrality point drops significantly and the quantization steps visible at higher angles are no longer appreciable. Furthermore, we consider the effects of a random distribution of vacancies on the quantum point contact. Whereas the electron-hole symmetry is broken in pristine samples, we find that this symmetry is restored upon increasing the concentration of vacancies. We explain this effect by a reduction of the effective interlayer coupling due to the presence of the vacancies.

\end{abstract}

\maketitle

\section{Introduction}

Quantum point contacts (QPC), short narrow conductors between two electron reservoirs that can be electrostatically defined or etched in two-dimensional materials, are the center of attention for their prospective applications in spin, valley and charge nanodevices~\cite{Sakanashi2021}. The goal is to control, by varying an external perturbation (typically an electrical potential), a degree of freedom of the system. Charge noise limits the coherence times compared to spin, so graphene spintronics has been an important area of research with the aim of achieving device applications~\cite{Han2014}. However, spin is hard to manipulate, so other degrees of freedom available in graphene-based devices, such as the valley index, are also the subject of current interest~\cite{Li2016}. 

In monolayer graphene, QPCs are created by etching, since a gate voltage does not open a gap in this material~\cite{Tombros2011}. However, etching generally produces defects which hinder the properties of the pristine material. In contradistinction, it is possible to open a gap in bilayer graphene with Bernal stacking by applying an electric field, so a great amount of effort has been devoted to elucidate and tune its transport properties via gate voltages. Furthermore, the fabrication of electrostatic QPCs in bilayer graphene has allowed to produce one-dimensional conducting channels and probe their transport behavior~\cite{Allen2012,Overweg2018NL}. QPC defined by electrostatic gating present fewer defects and have been shown to hold valley transport, the properties of which are still being understood~\cite{Li2016,Overweg2018PRL,Sakanashi2021}. 

Despite the aforementioned occurrence of defects and roughness in lithographically defined graphene QPCs, much effort has been devoted to improve these methods, and recently, graphene nanoconstrictions with high quality edges have been fabricated employing a cryo-etching technique~\cite{Caridad2018,Clerico2018}. Due to the low edge roughness and minimal defects, quantization of electron transport was experimentally observed, demonstrating the fabrication of low-defect QPCs via etching processes. This is especially important for the study of twisted bilayer graphene  (TBG) QPCs, in which the interplay of moir\'{e} and edge localization should be studied in systems with well-defined edges~\cite{SuarezMorell2014,SuarezMorell2015,Pelc2015}. In moir\'{e} systems, the interlayer coupling is inhomogeneous, being stronger in regions with direct or AA stacking than in AB-stacked parts. Consequently, edge states arising from AB-stacked regions are closer to the Fermi energy than those stemming from AA zones.  Such nontrivial interplay of edge states and moir\'{e} patterns has been evidenced in the identification of beadlike states of moir\'{e} edges in graphite with stacking disorder~\cite{SuarezMorell2014}. Hence, it is compelling to explore how such localization might be affected by vacancies, which are themselves the source of localized states. Additionally, issues such as the enhancement of strong correlations near the magic angle in one-dimensional moir\'{e} systems~\cite{ArroyoGascon2020,ArroyoGascon2023},  the nontrivial band topology of twisted bilayer graphene~\cite{Koshino2019}, the possible existence of gapped edge states related to its higher-order topology~\cite{Park2019} or even corner states~\cite{Park2021} in this material, have spurred the experimental exploration of quasi-one-dimensional moir\'{e} structures~\cite{Chen2021,FortinDeschenes2022,Wang2023}. In fact, twisted bilayer graphene nanoribbon (GNR) junctions have been experimentally shown to present highly tunable edge states with energy and spin degeneracy crucially dependent on the edge stacking offset~\cite{Wang2023}. Despite the progress on fabrication techniques, some defects or impurities might be present either due to the edge shaping process or to the adsorption of hydrogen in the GNR, so it is relevant to address the role of defects in these twisted moir\'{e} QPCs, in order to assess the change induced in their transport properties.

In this work, we study electron transport in QPCs based on TBG in the ballistic regime. In particular, we focus on the impact of single vacancies that can appear during growth and its effect on the transport properties of such QPCs. We analyze the interplay of vacancy-related localized states with edge states, which have been shown to be affected by the inhomogeneous coupling arising from the twist angle. In pristine samples, conductance is quantized for moderately large twist angle (larger than $\simeq 10\degree$), but quantized plateaus are hardly noticeable on approaching the magic angle ($\simeq 1.1\degree$). As it is well-known, the occurrence of moir\'{e} patterns breaks the electron-hole symmetry. However, we find that the electron-hole symmetry is partially restored on increasing the concentration of vacancies. We also explore the consequences of the spatial distribution of vacancies over the QPC, verifying that they have a larger impact when they are mainly located at the central part for systems with armchair edges.  

The paper is organized as follows. In Sec. II we present the system and model Hamiltonian used for obtaining electronic transport properties. In Sec.~III we present our results in two steps. First, in Sec.~III~A we investigate pristine QPCs and investigate the dependence of the conductance on the twist angle. Second, in Sec.~III~B we study the effect of a random distribution of single vacancies on the conductance. Section~IV concludes with a brief summary of the results.

\section{System and model Hamiltonian} \label{sec:Model}

\begin{figure}[ht]
    \centering
    \includegraphics[width=0.95\columnwidth]{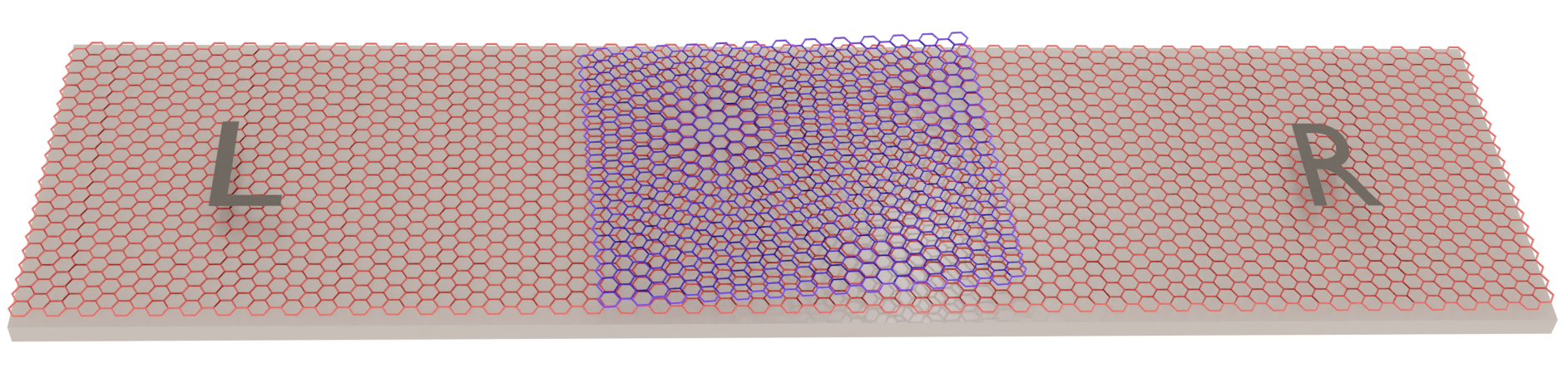}
    \caption{Sketch of the device that consists of a small graphene flake placed on top of a very long armchair GNR of the same width. The top flake is rotated with respect to the bottom GNR. The two regions of the long GNR away from the top flake are regarded as ideal leads, labelled L~(left) and R~(right).}
    \label{fig:setup}
\end{figure}

The system under consideration consists of a very long armchair GNR, on top of which a ultrasmall graphene flake of the same width is placed. The top flake is rotated with respect to the bottom GNR and it can be viewed as a graphene quantum dot. The two regions of the bottom GNR away from the top flake are regarded as ideal leads, denoted left~(L) and right~(R) in Fig.~\ref{fig:setup}. Therefore, electrons coming from the left lead are scattered off from the middle region that is coupled to the top flake and then they are reflected or transmitted to the right lead. In this way a QPC is created.



The tight-binding Hamiltonian reads as $H_1+H_2+H_\mathrm{inter}$, where $H_{\ell}$ describes the electron dynamics in the lower~($\ell=1$) and upper~($\ell=2$) graphene monolayers (the ribbon and the flake, respectively) within the nearest-neighbor approximation and $H_\mathrm{inter}$ takes into account the interlayer coupling. Hence
\begin{subequations}
\begin{align} 
   \mathcal{H}_{\ell}&=-\sum_{\langle i,j\rangle}\gamma_{0}\,c^\dagger_{\ell i}\,c_{\ell j}^{}+\mathrm{H.c.}\ , 
   \label{eq:01a} \\ 
   \mathcal{H}_\mathrm{inter}&=-\sum_{i,j}\gamma_{1}\,e^{-\beta({r}_{ij}-d)}\,c^\dagger_{1i}\,c_{2j}^{}+\mathrm{H.c.}\ ,
   \label{eq:01b}
\end{align}
\label{eq:01}%
\end{subequations}
where $\mathrm{H.c.}$ stands for Hermitian conjugate and the summation in $\langle i,j\rangle$ runs over nearest-neighbor C atoms. The origin of energy is set at the energy of the C orbital. Here $c^{\dagger}_{\ell i}$ and $c_{\ell i}^{}$ are the creation and annihilation fermion operators at site $i$ of the monolayer $\ell$. The inter- and intralayer tunnel energies are $\gamma_{0}=3.16$\;eV and $\gamma_{1}=0.39$\;eV, respectively; $r_{ij}=|{\bm r}_i-{\bm r}_j|$ is the distance between atoms with positions ${\bm r}_i$ and ${\bm r}_j$, each one located on a different layer; $d=3.35\,$\AA\ is the interlayer distance and $\beta=3\,$\AA$^{-1}$~\cite{SuarezMorell2014}. The summation in~\eqref{eq:01b} runs over all atom pairs $i$ and $j$ of different layers, and it is restricted to $r_{ij}<6a_0$, with $a_0=1.42\,$\AA\ being the C--C distance. This parameterization has been proved to provide excellent agreement with ab-initio calculations of the bands near the Fermi energy of TBG systems~\cite{FlatBands,ArroyoGascon2020,ArroyoGascon2023}.

Electron states of the QPC are characterized by the local density of states (LDOS), defined as
\begin{subequations}
\begin{equation}
    \rho({\bm r}_i,E)=\sum_{\nu} \big| \psi_\nu({\bm r}_i) \big|^2\delta\left(E-E_\nu\right)\ ,
    \label{eq:02a}
\end{equation}
where the summation runs over all eigenstates of the QPC with energy $E_\nu$ and $\psi_\nu({\bm r}_i)$ denotes the amplitude at position ${\bm r}_i$ of the corresponding eigenstate. The wave function is assumed to be normalized hereafter. Similarly, the density of states (DOS) is calculated from the LDOS as $\rho(E)=\sum_{i=1}^{N} \rho({\bm r}_i,E)$, where $N$ is the number of atomic positions in the QPC.

The spatial extent of an arbitrary eigenstate $\nu$ can also be estimated from the participation ratio ($\mathrm{PR}$)
\begin{equation}
    \mathrm{PR}_\nu=\left(\sum_{i=1}^{N}\big| \psi_\nu({\bm r}_i) \big|^4\right)^{-1}\ .
    \label{eq:03b}
\end{equation}
Recall that $\mathrm{PR}_\nu\sim 1$ when the eigenstate is localized at a single atom and $\mathrm{PR}_\nu\sim N$ when the eigenstate is extended over the whole QPC. Finally, electron transport properties at low temperature were calculated within the framework of the Landauer--Büttiker formalism~\cite{datta}. In this formalism, the conductance is calculated in the linear regime (i.e., at low bias) as 
\begin{equation}
    G(E)=\frac{e^2}{h}\,\tau(E)\ ,
    \label{eq:03c}
\end{equation}
\label{eq:03}%
\end{subequations}
where $\tau (E)$ is the transmission coefficient at energy $E$. Electrons scatter off the QPC and conductance is found to be dependent on the number of modes which can travel along the device. Calculation of the three magnitudes~\eqref{eq:03} was performed with the help of the {\tt Kwant} toolkit~\cite{kwant}. 

\section{Results} \label{sec:Results}

\begin{figure*}[ht]
    \centering
\includegraphics[width=2.0\columnwidth]{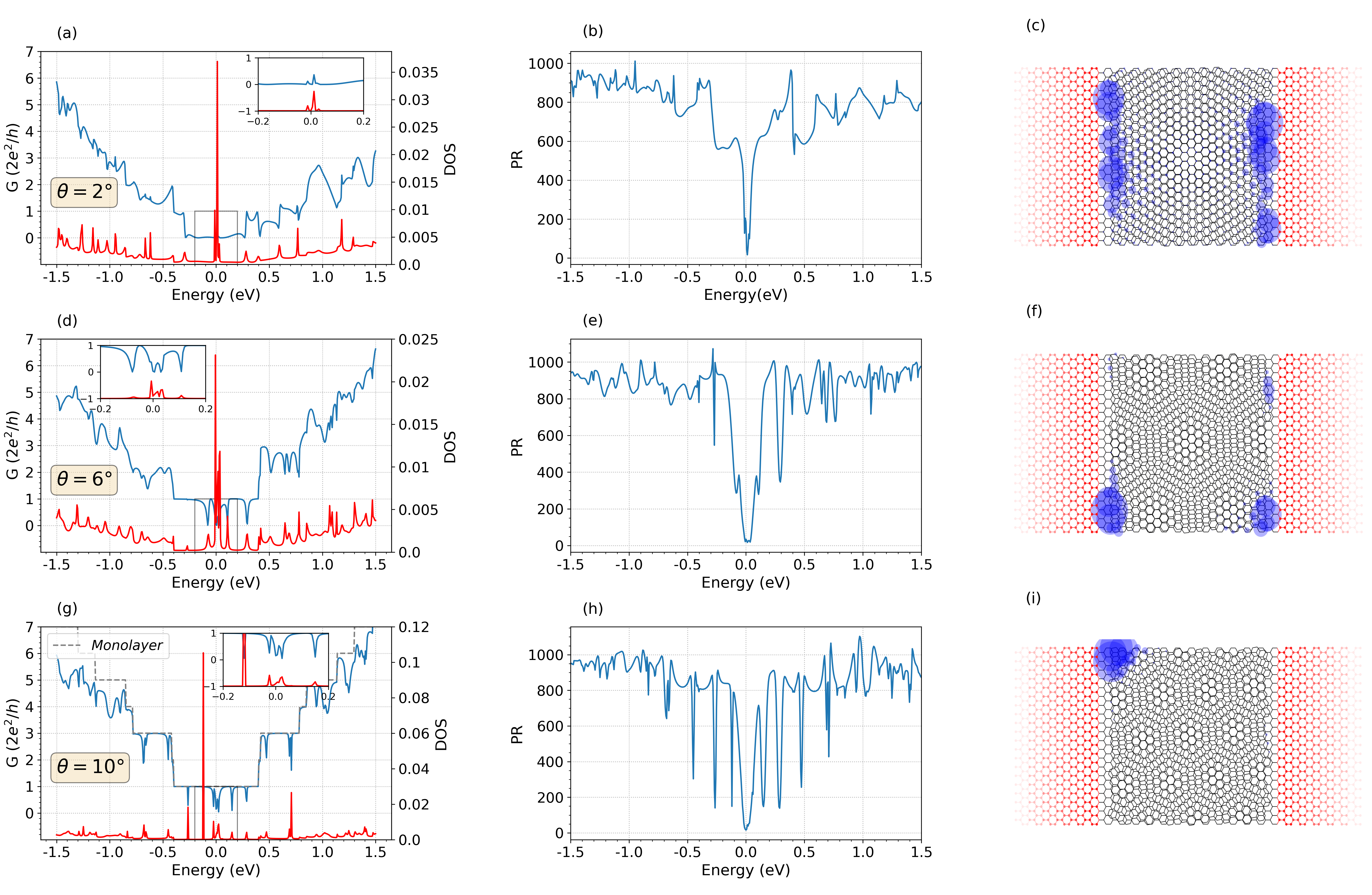}
    \caption{Leftmost panels show the conductance (blue line) and DOS of the QPC (red line) for rotation angles of $\theta=2\degree$ (a), $\theta=6\degree$ (d) and $\theta=10\degree$ (g). Middle panels show the $\mathrm{PR}$ for $\theta=2\degree$ (b), $\theta=6\degree$ (e) and $\theta=10\degree$ (h). The rightmost panels represent the LDOS at energy $E=0$ for each rotation angle. The radii of the circles are proportional to the LDOS on each atom. Left and right leads appear in red color.}
    \label{FIGURA2}
\end{figure*}

Our results were obtained for a QPC of dimensions $5\times5$\,nm$^2$, which contains of the order of $1900$ atoms. This size suffices to observe the moir\'{e} pattern at moderately small angles and does not demand excessive computational cost. Due to the electron-hole symmetry breaking, we consider both positive and negative energies in the range $\pm 1.5$\;{\rm eV} around the charge neutrality point. 

\subsection{Pristine QPC}

First we consider a pristine QPC to assess the electron dynamics in the absence of vacancies. Figure~\ref{FIGURA2} depicts the conductance, $\mathrm{PR}$ and LDOS as functions of the energy for rotation angles $\theta$ of $2\degree$, $6\degree$, and $10\degree$. The conductance curves are asymmetrical around the charge neutrality point, presenting higher values at negative energies for $\theta=2\degree$, but at positive energies in the case of $\theta=6\degree$ and $\theta=10\degree$. This electron-hole broken symmetry is related to the extent of the interlayer coupling, that reaches all neighboring atoms within a circle of 8.52 \AA\ radius. As discussed in the context of TBG, this coupling mixes the graphene sublattices,  breaking electron-hole symmetry. Furthermore, the conductance presents a marked dependence with the rotation angle. At $\theta=10\degree$ the three plateaus of the quantized conductance are perfectly observable up to $\pm 0.8$ eV, and coincide with those of a single GNR. On these plateaus, clear antiresonances stemming from the localized states of the top flake are distinguishable and identified by the coincidence of the DOS peaks, as it can be verified in Fig.~\ref{FIGURA2}(g). On decreasing the angle the moir\'{e} periodicity is less apparent since the flake size becomes smaller than the moir\'{e} period, and the plateaus of the conductance are smeared out, being barely noticeable for $\theta=2\degree$. Figs. 1 and 2 in the Supplemental Material support this assertion with more rotation angles.

\begin{figure*}[ht]
    \centering
    \includegraphics[width=2.0\columnwidth]{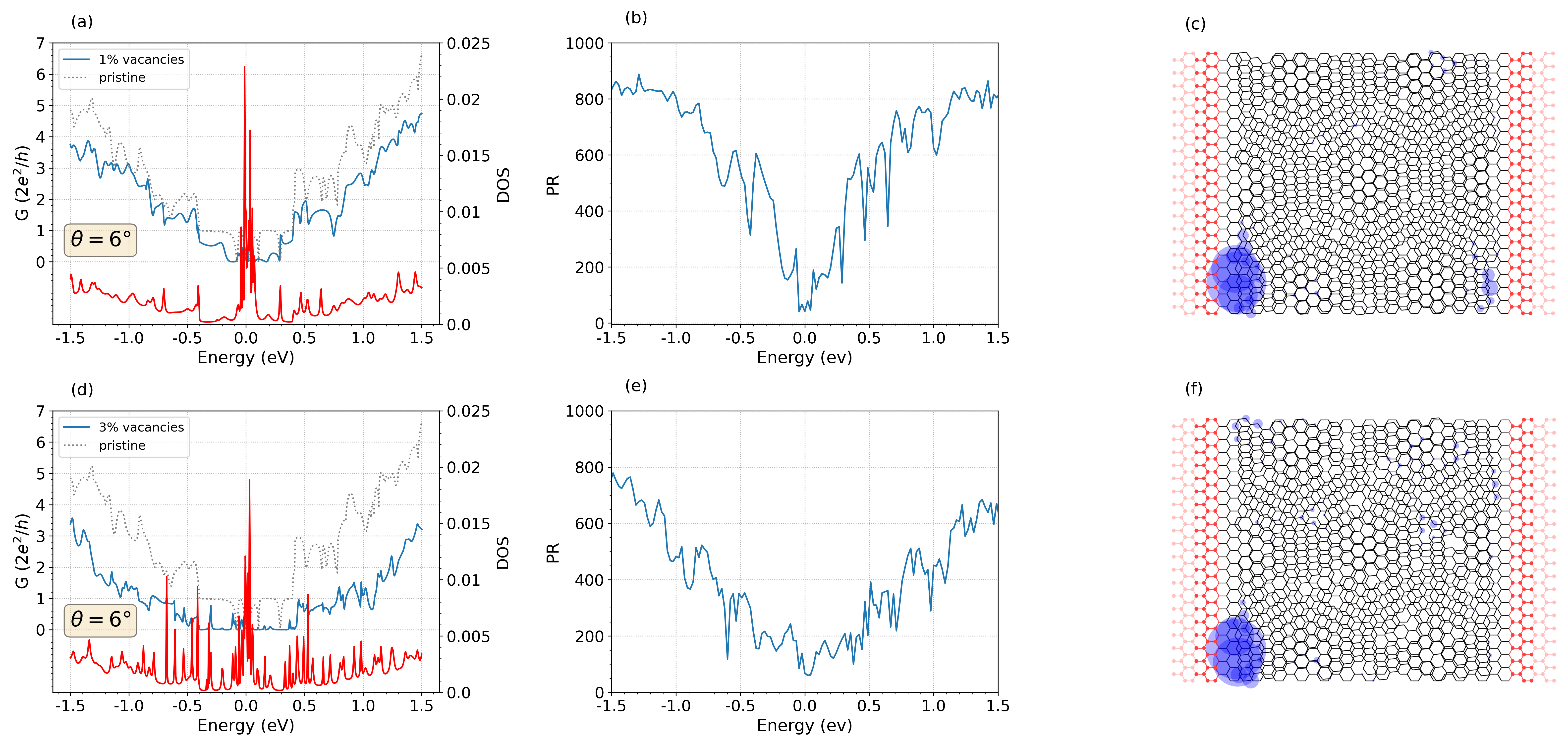}
    \caption{Leftmost panels show conductance of a pristine QPC (grey dashed line) and a defective QPC (blue line) at $\theta=6\degree$. In the latter case, vacancies are randomly distributed over the whole QPC with concentration (a)~$1\%$ and (d)~$3\%$. The DOS of the QPC is also displayed (red line). Middle panels show the $\mathrm{PR}$ for a vacancy concentration of (b)~$1\%$ and~(e) $3\%$. The rightmost panels show the LDOS at $E=0$ for each concentration. The radii of the circles are proportional to the LDOS  on each atom.}
    \label{FIGURA3}
\end{figure*}
 
The DOS, shown in Figs.~\ref{FIGURA2}(a), \ref{FIGURA2}(d) and \ref{FIGURA2}(g), provides information of the states of the leads as well as the discrete states of the QPC. The DOS presents sharp peaks due to the occurrence of discrete levels that result from quantum confinement effects in the QPC, which can be viewed as a coupled quantum dot to a GNR conductor. These spatially-localized states interfere with the continuum states of the leads, giving rise to Fano-like resonances in the conductance. This behavior was also observed in bilayer flakes with AA or AB stacking, for which these destructive interferences were also analyzed~\cite{gonzalez2010electronic,Gonzalez2011}. The peak of the conductance at $E=0$ for all rotation angles is clearly seen in Fig.~\ref{FIGURA2}. The states associated to these peaks are highly localized, in accordance with the small value of the $\mathrm{PR}$ observed in Fig.~\ref{FIGURA2}. The inset of the left panels of Fig.~\ref{FIGURA2} show an enlarged view of the conductance close to the charge neutrality point. It is found that there are two very close peaks at positive and negative energy. The spatial extent of the corresponding states can be revealed by the atom-resolved LDOS, presented in the rightmost panels of Fig.~\ref{FIGURA2}. The states are edge-like, localized at the transverse edge of the QPC, perpendicularly to the direction of the current. In the case of $\theta=2\degree$, the states spread along the edge, while in the other two angles they are corner-like states. This can be explained in view of the zigzag character of the edge that causes the appearance of edge states. The transverse edge is mainly zigzag for $\theta=2\degree$, but on increasing the angle, the edges becomes more chiral (i.e., larger mixture of zigzag and armchair portions) and prevent the spreading of the states. Also, it is worth mentioning that the two edge states for $\theta=2\degree$ decay inside the QPC [Fig.~\ref{FIGURA2}(c)], but are weakly coupled due to the small size of the QPC. This coupling is responsible of the small shift from the charge neutrality point of the two peaks, which hybridize to form bonding and antibonding states. For $|E|>0.4\,$eV the DOS gradually grows as the number of occupied modes of the leads increases.

\subsection{QPC with vacancies}

We now turn the attention to the effects of point vacancies located in the QPC on electron transport properties. These defects are introduced by removing atomic sites and the corresponding bonds in both monolayers of graphene. First, we study a random distribution of vacancies over the whole QPC (vacancies are not included in the leads). Next, we restrict the location of the vacancies to the edges and to the middle region of the QPC to elucidate their impact on the conductance. 

\begin{figure*}[ht]
    \centering
\includegraphics[width=2.0\columnwidth]{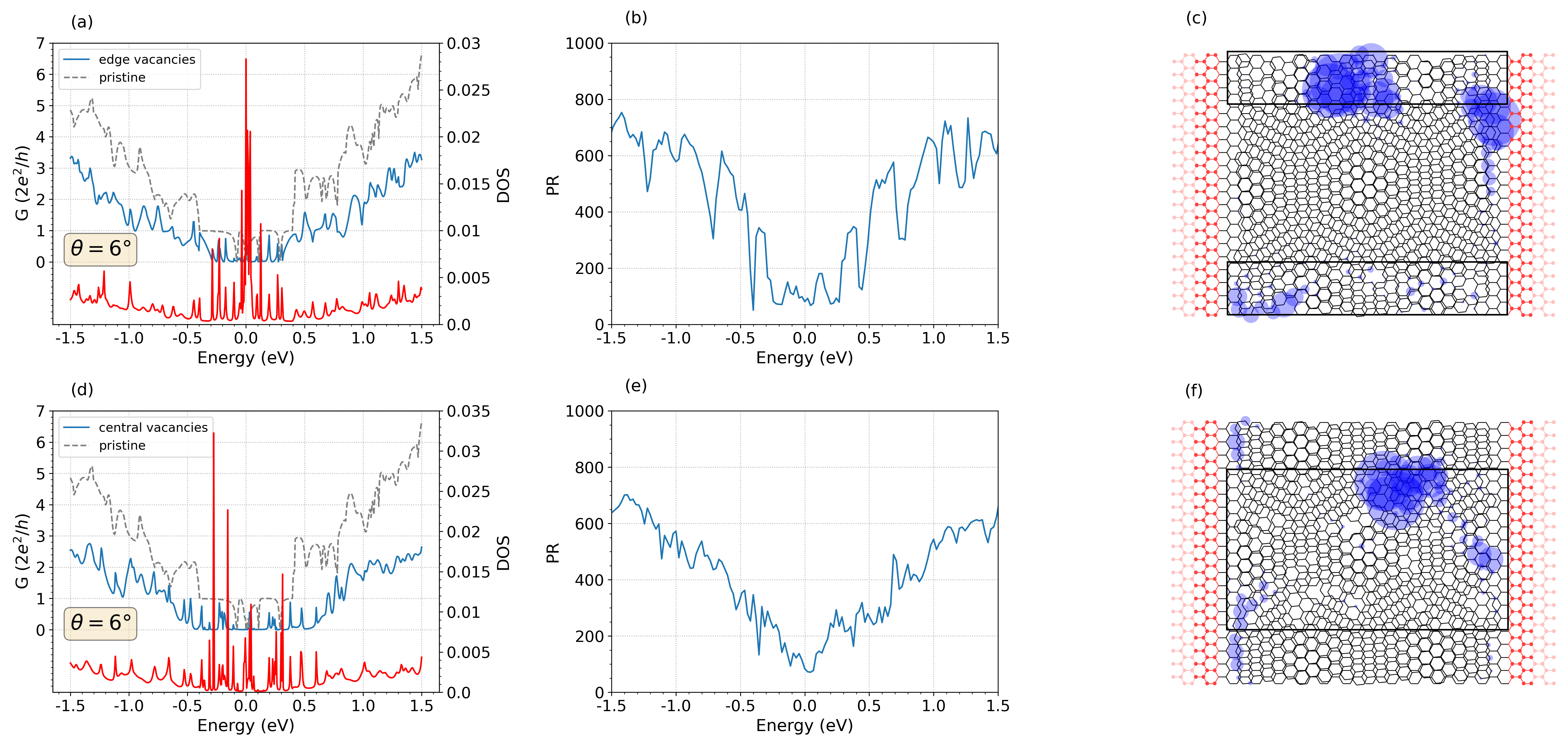}
    \caption{Conductance as a function of energy of a pristine QPC (grey dashed line) and a defective QPC in the case of $\theta=6\degree$ when the vacancies are located at~(a) edges and~(d) central part of the QPC. Panels~(b) and~(e) show the $\mathrm{PR}$ corresponding to panels~(a) and~(d), respectively. Panels~(c) and~(f) display the LDOS at energy E=0 for each case. The radii of the circles are proportional to the LDOS on each atom. The concentration of vacancies is $3\%$ in both cases.}
    \label{FIGURA4}
\end{figure*}

Transport properties in a nanometric system are found to depend noticeably on the particular arrangement of vacancies of each disorder realization. Hence, an average over different disorder realizations does not provide useful information, as it is usual for nanoscopic systems. Instead, we choose some particular configurations to present the typical electrical response of the QPC. Taking the intermediate angle $\theta=6\degree$ of the pristine QPC (Fig.~\ref{FIGURA2}) as an example of twisted angle, we show in Fig.~\ref{FIGURA3} the effect of adding a distribution of vacancies covering at random the whole area of the QPC, for concentrations of $1\%$ and $3\%$ relative to the total QPC sites. The $3\%$  case was generated from the configuration with $1\%$ and adding extra vacancies, so that both concentrations share part of the vacancy positions. A first result observed in Figs.~\ref{FIGURA3}(a) and~\ref{FIGURA3}(d) is the  drop of the conductance and the suppression of the previously smeared out  plateaus (with decreasing twist angle) due to scattering. This effect raises with the vacancy concentration. It is also noticeable a gap in the conductance that appears at low energies, which broadens on increasing the concentration. Vacancies in graphene are known to induce localized states with energies close to the Dirac point \cite{Ma2004,Pereira96}. These localized states interfere with the continuum of states as Fano resonances, causing zero-conductance dips. These dips were later extensively studied in graphene nanoribbons~\cite{PRB77,conductionsupression,PRB88}, where the shape and energy of the dip depend on the geometry of the edge and the position of the vacancy. However, when the edge is armchair the dip remains around $E=0$ for both monolayer and bilayer ribbons. In the case of this work, where there are many vacancies, the single vacancy antiresonances add up, leading to the formation of a transport gap around zero energy~\cite{PRB88}. The DOS confirms this affirmation,  with many peaks flocking together around this energy due to the appearance of  vacancy states, as expected~\cite{Ma2004,PRB77}, in contrast to the pristine case presented in Fig.~\ref{FIGURA2}.

But the most intriguing effect caused by vacancies is that the conductance curves become more symmetrical than in the pristine case, and this symmetry turns out to increase with the vacancy concentration [see Figs.~\ref{FIGURA3}(a) and~\ref{FIGURA3}(d)]. This phenomenon not only occurs for the chosen arrangement of vacancies but in many other studied configurations, as shown in Fig. 3 in the Supplemental Material. Therefore, we find a recovery of the electron-hole symmetry, clearly broken in twisted bilayer graphene, with the inclusion of vacancies. Broken symmetry for this model of TBG occurs due to the large span of the interlayer hopping which, unlike the usual first-neighbor hopping models in monolayer graphene, enables tunnel processes between atoms of different sublattices. This results in a sublattice mixing which causes electron-hole symmetry breaking. We claim that the addition of vacancies weakens the mixing of sublattices  because it reduces the number of interlayer tunnel processes, so that the electron-hole symmetry is partially restored.

The $\mathrm{PR}$ of electron states is also affected by vacancies. Compared to the pristine case [see Fig.~\ref{FIGURA2}(e)], the region with low $\mathrm{PR}$ around the charge neutrality point of energy becomes broader on adding vacancies, as it can be observed in Figs.~\ref{FIGURA3}(b) and~\ref{FIGURA3}(e). This means that states are more localized in this case, according with the previous analysis. Finally, we explore the energy $E=0$ and plot the LDOS in Figs.~\ref{FIGURA3}(c) and~\ref{FIGURA3}(f). The localization of the $E=0$ state around the vacancies  can be noticed under a close inspection. However, localization is stronger at the zigzag edges of the top flake; in fact, the $E=0$ state is mainly located at the edges of the QPC as in the pristine QPC [see Fig.~\ref{FIGURA2}(f)].

In addition, we study the dependence of transport properties with the relative position of vacancies inside the QPC, in a similar way to Ref.~\cite{PRBstrongdisorder}. For that purpose, we study two different spatial configurations of vacancies located at the edge and central regions. In the first configuration, we add vacancies randomly but restricted inside two areas at the longitudinal edges of the QPC, i.e., those parallel to the current direction. Each area covers $1\times 5\,$nm$^{2}$ [see the areas in Fig.~\ref{FIGURA4}(c)]. In the second configuration, we restrict the random vacancies only to the middle region of the QPC, in an area of $3\times 5\,$nm$^{2}$ [see the area in Fig.~\ref{FIGURA4}(f)]. The sum of the areas with possible vacancies of the first and second configurations gives the entire flake area. The vacancy concentration is the same for both configurations, namely $3\%$ relative to the total QPC sites. It is observed that for the edge geometry studied in this work, with armchair edges in the direction of the current, vacancies located at the edges are less detrimental for electron transport, producing a smaller drop in the conductance than the vacancies placed in the middle region [Figs.~\ref{FIGURA4}(a) and~\ref{FIGURA4}(d)]. This can be explained by analyzing the spatial distribution of the electronic wave function in the flake. The amplitude of the wave function of an armchair system practically vanishes at the edges, which causes the overlap with localized vacancy states to be smaller when vacancies are located at the edges. Therefore, the detrimental effect on the conductance is less compared to the case of  vacancies localized at the central part of the flake. This also explains the smaller conduction gap at the edge configuration. In both configurations the number of vacancies is equal to the case of the distribution spread over the whole QPC of the Fig.~\ref{FIGURA3}(f). However, as the defects are restricted to smaller regions in the configuration with vacancies at the edges, the intervacancy distance is lower. Being the vacancies closer, their local densities add up, yielding a larger LDOS in Figs.~\ref{FIGURA4}(c) and \ref{FIGURA4}(f). Notice that in this case the LDOS  is not dominated by localization at the edges, as it was the situation depicted in Fig.~\ref{FIGURA3}(f).

\section{Conclusions}\label{sec:concl}

In summary, we have studied numerically electron transport through QPCs made of twisted bilayer GNRs at low temperature in the linear regime. The twist angle determines the electronic conductance of pristine samples. On one side, decreasing the angle to approach the magic angle ($\theta \sim 1.1\degree$) makes the quantized plateaus hardly observable. On the other side, the occurrence of moir\'{e} patterns break the electron-hole symmetry, as revealed by the conductance curves when the Fermi level is shifted from the charge neutrality point. We have also considered fabrication imperfections in the form of single vacancies randomly distributed over the QPC. We studied three different arrangements of the vacancies, namely, uniformly spread over the whole QPC, located in the central region of the QPC or at the edges of the QPC. Remarkably, the electron-hole symmetry is partially recovered upon increasing the concentration of vacancies that are uniformly distributed over the QPC. The recovery can be attributed to the reduction of the effective interlayer coupling. Finally, the relevance of edge states in the electron transport is determined from the smaller detrimental effect of the conductance when the vacancies are located at the edges.

\acknowledgments

This work was supported by the Spanish Ministry of Science and Innovation (grant PID2019-106820RB-C21). We also acknowledge the support from the “(MAD2D-CM)-UCM” project funded by Comunidad de Madrid, by the Recovery, Transformation and Resilience Plan, and by NextGenerationEU from the European Union.

%


\end{document}